\documentclass[english,english,aip,jcp,graphicx,twocolumn,showpacs]{revtex4}
\usepackage{graphicx}

\makeatletter

\makeatletter

\usepackage{subfigure}

\makeatother

\usepackage{babel}
\makeatother

\begin{document}

\title{Fractional Spins and Static Correlation Error in Density Functional Theory}

\author{Aron J. Cohen, Paula Mori-S\'anchez and Weitao Yang}

\affiliation{Department of Chemistry, Duke University, Durham, North Carolina
27708}


\date{\today{}}

\begin{abstract}
Electronic states with fractional spins arise in systems with large
static correlation (strongly correlated systems). Such fractional-spin
states are shown to be ensembles of degenerate ground states with
normal spins. It is proven here that the energy of the exact functional
for fractional-spin states is a constant, equal to the energy of the
comprising degenerate pure spin states. Dramatic deviations from
this exact constancy condition exist with all approximate functionals,
leading to large static correlation errors for strongly correlated
systems, such as chemical bond dissociation and band structure of
Mott insulators. This is demonstrated with numerical calculations for
several molecular systems. Approximating the constancy behavior for
fractional spins should be a major aim in functional constructions
and should open the frontier for DFT to describe strongly correlated
systems. 
The key results are also shown to apply in reduced density-matrix functional theory.
\end{abstract}
\maketitle
Density functional theory (DFT) \cite{Kohn651133,Parr89} is a rigorous
approach for describing the ground state of any electronic system.
The success or failure of DFT is based on the quality of the density
functional approximation (DFA). One of the dramatic failures is in
strongly correlated systems, characterized by the presence of degeneracy
or near degeneracy \cite{Savin96327}, having large static correlation.
The simplest example is the dissociation of H$_{2}$ molecule \cite{Baerends01133004,Becke05064101} 
for which commonly used DFAs over-estimate the energy by more than 50 kcal/mol. Closedly related
are the band structure of Mott insulators \cite{Fazekas99} which are described as metallic
by known DFAs and problems in describing superconducting cuprates \cite{Perry002438}.

The improvement of the DFA is, therefore, a major goal that critically depends
on the underlying theoretical construction. One of the most useful developments is the extension 
of DFT to fractional charges developed by Perdew \textit{et. al.} \cite{Perdew821691}
in a grand canonical ensemble, which was also established later in
a pure state formulation \cite{Yang005172}. For a system with fractional
charges, the exact energy is a straight line interpolating the energies
of the integer electron systems. The violation of this exact condition 
leads to two types of errors \cite{Morisanchez08146401}, the delocalization error (DE) of most functionals like LDA,
GGA and hybrids  \cite{Becke883098,Lee88785,Perdew963865,Becke935648}, and the localization error (LE) 
of the Hartree-Fock functional.
DE captures the tendency of commonly used functionals to bias toward
a delocalized description of electrons
with widespread implications from molecular reactions\cite{Morisanchez06201102} to the band-gap of solids.
Addressing this error resulted in the construction of the MCY3 and rCAMB3LYP functionals \cite{Cohen07191109},
that correct many of the errors of previous functionals. In particular
they correctly predict the ionization energy and electron affinity from
their single-electron eigenvalues, and hence the energy gaps in molecules
\cite{Cohen08115123}.

In this Letter, we make an extension of DFT to fractional spin systems
and prove that the exact energy functional of a fractional spin state
is that of the comprising degenerate normal spin states. We show that
states with fractional spins arise in systems with large static correlation
(strongly correlated systems) and that the dramatic deviation from
the proven exact condition accounts for large static correlation errors.
We also introduce a quantitative measure for static correlation error
(SCE). 

Our starting point is the exact result for an ensemble of degenerate
densities derived by Yang, Zhang and Ayers (YZA) \cite{Yang005172}:
For a $N$-electron system in the external potential $v({\bf r})$
that has $g$-fold degenerate orthogonal ground state wavefunctions
$(\Phi_{i},i=1,2,\ldots,g)$ with corresponding densities $(\rho_{i},i=1,2,\ldots,g)$
and ground state energy $E_{v}^{0}(N)$, the ensemble density is $\rho=\sum_{i=1}^{g}C_{i}\rho_{i},$
where $0\leq C_{i}\leq1$ and $\sum_{i=1}^{g}C_{i}=1.$ The exact
energy functional satisfies the following equation \begin{equation}
E_{v}\left[\sum_{i=1}^{g}C_{i}\rho_{i}\right]=E_{v}\left[\rho_{i}\right]=E_{v}^{0}(N),\label{Theorem}\end{equation}
 if $E_{v}^{0}(N)\leq(E_{v}^{0}(N+1)+E_{v}^{0}(N-1))/2.$ 
Note that in the derivation of Eq. (\ref{Theorem}) only pure states were
used and the ensemble densities appear in the limit of large separation
of fragments \cite{Yang005172}.


We now examine the application of Eq. (\ref{Theorem}) to fractional
spin systems. Consider a $N$-electron atom or molecule that is a
doublet, with total spin $S=\frac{1}{2}$, for example, the H atom
with $N=1$. It has two degenerate spin states labeled with the $z$-component
of the spin $m_{s}=\frac{1}{2}$ and $m_{s}=-\frac{1}{2}$. Now we
construct $\rho_{{\rm fs}}(S,\gamma)$, an ensemble density with fractional
spins as 
\begin{equation}
\rho_{{\rm fs}}\left({\textstyle \frac{1}{2}},\gamma\right)=\left(\frac{1}{2}+\gamma\right)\rho({\textstyle \frac{1}{2},\frac{1}{2}})+\left(\frac{1}{2}-\gamma\right)\rho({\textstyle \frac{1}{2},-\frac{1}{2}}),\label{eq:Hatom}\end{equation}
 where $\rho(S,m_{s})$ is the ground state density with $m_{s}$
and $\gamma$ $(-\frac{1}{2}\leq\gamma\leq\frac{1}{2})$ 
is the net $z$-component of the spin in the fractional-spin state.
$\rho_{{\rm fs}}(S,\gamma)$ represents many fractional spin states. In
particular, $\gamma=0$ represents a state that has half a spin-up
electron and half a spin-down electron occupying the same spatial
orbital, its total spin density being equal to zero everywhere. Applying
the YZA relation (Eq. (\ref{Theorem})) leads to \begin{equation}
E_{v}\left[\rho_{{\rm fs}}\left({\textstyle {\frac{1}{2}},\gamma}\right)\right]=E_{v}\left[\rho({\textstyle {\frac{1}{2}},{\textstyle {\frac{1}{2}})}}\right]=E_{v}\left[\rho({\textstyle {\frac{1}{2}},-{\textstyle {\frac{1}{2}})}}\right],\label{eq:FracE1}\end{equation}
 which shows that for the exact density functional, all the fractional-spin
densities $\rho_{{\rm fs}}\left(\frac{1}{2},\gamma\right)$ have the same
degenerate energy. However, all known DFAs fail
dramatically and give much too high an energy for $E_{v}\left[\rho_{{\rm fs}}\left(\frac{1}{2},\gamma\right)\right]$,
as illustrated in the right-hand side of Fig. 1. In the self-consistent
spin-unrestricted Kohn-Sham (KS) calculations, we use the spin densities
$\rho_{{\rm fs}}^{\sigma}(S,\gamma)$ $(\sigma=\alpha,\beta)$, which are
represented by a noninteracting system with fractional occupation
numbers, \begin{equation}
\rho_{{\rm fs}}^{\sigma}(S,\gamma)=\sum_{i}^{{\rm HOMO}}n_{i}^{\sigma}\left|\phi_{i}^{\sigma}\right|^{2},\label{eq:KSdensity}\end{equation}
 where only the highest occupied molecular orbital (HOMO) for each spin is
fractionally occupied, with $n_{\rm HOMO}^{\alpha}=\frac{1}{2}+\gamma$
and $n_{\rm HOMO}^{\beta}=\frac{1}{2}-\gamma.$ At $\gamma=0$, the ground
state is spin-unpolarized with $\rho_{{\rm fs}}^{\alpha}(\frac{1}{2},0)=\rho_{{\rm fs}}^{\beta}(\frac{1}{2},0)$
everywhere, and the deviation of the constancy requirement Eq. (\ref{eq:FracE1})
reaches its maximum for DFAs.

In carrying out KS calculations with Eq. (\ref{eq:KSdensity}),
we have used the following variational principle 
\begin{equation}
E_{v}\left[\rho_{{\rm fs}}\right]=\min_{\tilde{\rho}_{{\rm fs}}}E_{v}\left[\tilde{\rho}_{{\rm fs}}\right],\label{eq:variational}\end{equation}
 where the domain of variation for $\tilde{\rho}_{{\rm fs}}$ is all
the ensemble densities constructed from Eq. (\ref{eq:Hatom}) 
but with arbitrary spin densities $\rho({\textstyle \frac{1}{2},\frac{1}{2}})$
and $\rho({\textstyle \frac{1}{2},-\frac{1}{2}})$. We have also assumed
that such ensemble densities can be represented by a noninteracting
system with fractional occupation. Details of the proof for Eq. (\ref{eq:variational})
can be found in Ref. \cite{Cohen08PRLSup}.

\begin{figure}[!t]
\includegraphics[width=0.5\textwidth]{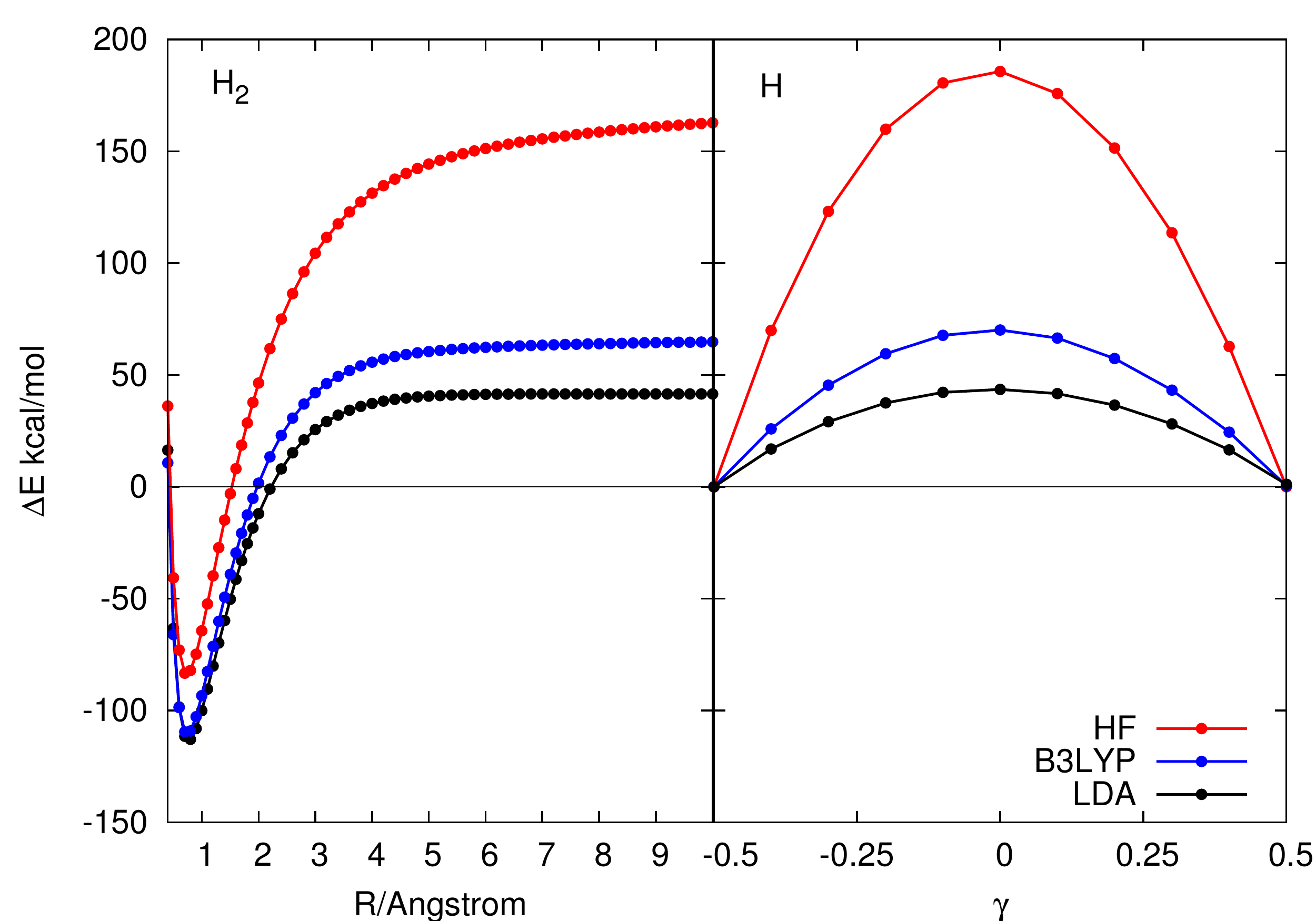} 
\caption{\label{fig1} Binding curve of H$_{2}$ calculated with spin-restricted
KS and fractional spins of H atom calculated with spin-unrestricted
KS (multiplied by 2). $\gamma=0$ is a H atom with half an $\alpha$
electron and half a $\beta$ electron, which is the dissocation limit
of H$_{2}$. All calculations are self-consistent using a cc-pVQZ
basis set.}
\end{figure}

This formalism is particularly interesting because the fractional spin
density $\rho_{{\rm fs}}^{\sigma}(\frac{1}{2},0)$ describes the dissociation
limit of a single chemical bond. For example, at the dissociation
limit of the H$_{2}$ molecule, a singlet system ($S=0$) is obtained,
which consists of two fractional spin H atoms separated by a large
distance. This system can be properly described by multi-configurational
wave function methods. However in DFT, spin-restricted KS calculations
having the correct spin state ($S=0$), give much too high an energy,
with DFAs. The over-estimation in the energy
for the dissociation of H$_{2}$ matches exactly the over-estimation
for the H atom with fractional spin density $\rho_{{\rm fs}}^{\sigma}(\frac{1}{2},0)$.

Fig. 1 illustrates the performance of three commonly used functionals,
LDA, B3LYP and Hartree-Fock (HF). The left-hand side shows the spin-restricted
binding curve of the H$_{2}$ molecule from the ground state unrestricted
atoms (with integer spins i.e. one alpha electron and zero beta electrons
or vice versa, corresponding to $\rho_{{\rm fs}}^{\sigma}(\frac{1}{2},\pm\frac{1}{2})$)
and the right-hand side shows the difference in energy of the H atom
with fractional spins, $\rho_{{\rm fs}}^{\sigma}(\frac{1}{2},\gamma$),
from the energy of the same ground state unrestricted atom (multiplied
by two for direct comparsion with the binding curve). 
The energy should be constant with the change in $\gamma$
but all the energy functionals have a very large error,
ranging from 30 kcal/mol to 170 kcal/mol for
the midpoint, $\gamma=0$. HF has the largest error and LDA has the
smallest error, but both functionals over-estimate the energy for
fractional spin states. B3LYP, as expected, has a behavior inbetween LDA and HF. 
Other functionals also suffer from large errors \cite{Cohen08PRLSup}, with
GGA functionals performing roughly the same as LDA and other hybrid functionals
somewhere inbetween LDA and HF (this also includes coulomb attenuated
functionals with reduced DE).
This suggests that the calculation of strongly correlated systems, where
this error is important, will qualitatively fail if any of the above functionals are used. 
There are many attempts in the literature to circumvent this error, 
for example breaking the spin symmetry, which gives reasonable energies but wrong spin densities.

%
\begin{figure}[!t]
\includegraphics[width=0.5\textwidth]{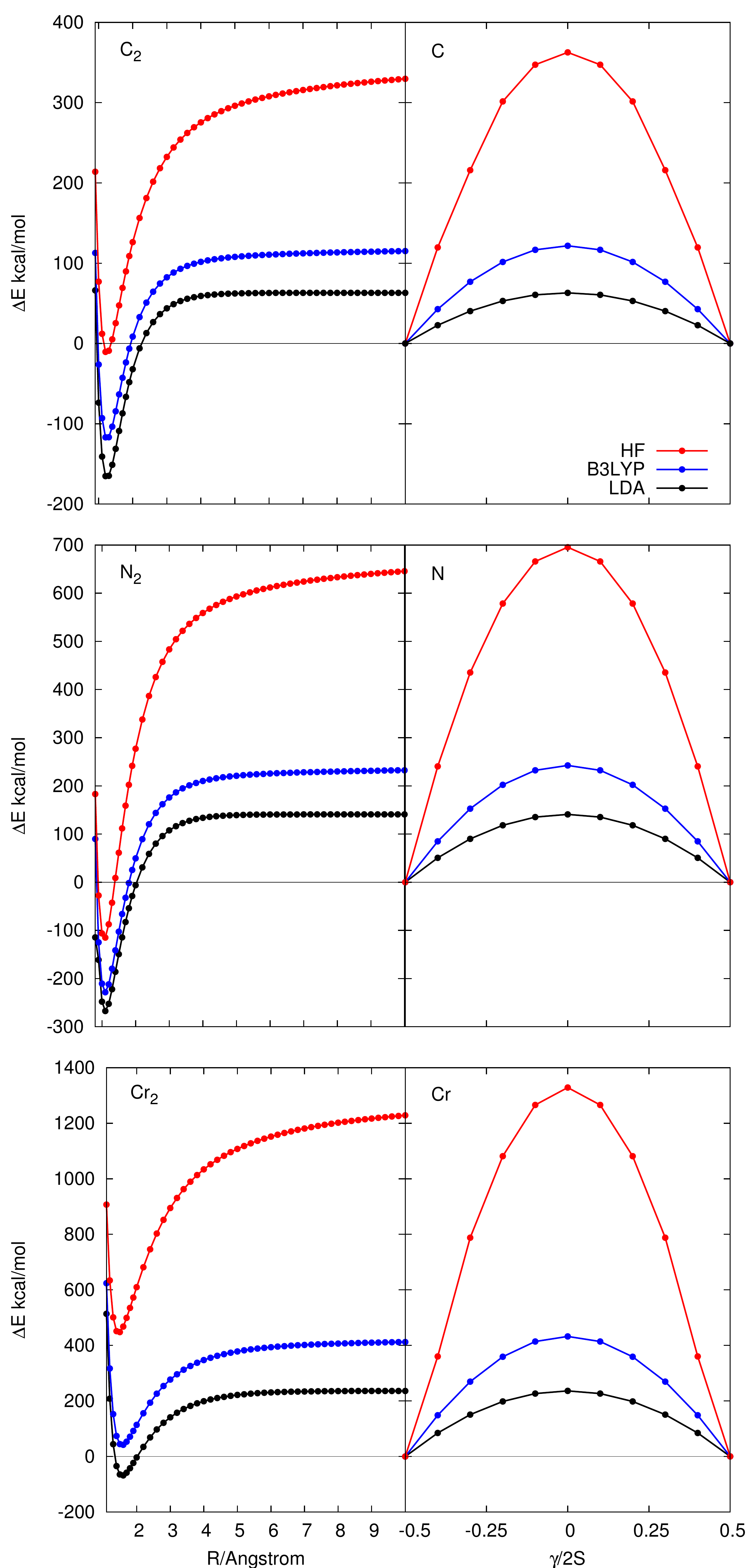} 

\caption{\label{fig2} The same as Fig. 1 but for top, C$_{2}$ (double bond),
middle, N$_{2}$ (triple bond) and bottom, Cr$_{2}$ (sextuple bond).}

\end{figure}

Our discussion for the single bond dissociation can be extended to
multiple bond dissociation. Using the notation in Eq. (\ref{eq:Hatom}),
for a system with total spin $S$, we can construct the fractional
spin density $\rho_{{\rm fs}}(S,\gamma)$ from the two degenerate spin states
with maximum $\left|m_{s}\right|=S$; \begin{equation}
\rho_{{\rm fs}}(S,\gamma)=\left(\frac{1}{2}+\frac{\gamma}{2S}\right)\rho(S,S)+\left(\frac{1}{2}-\frac{\gamma}{2S}\right)\rho(S,-S),\label{eq:OtherAtom}\end{equation}
 where $-S\leq\gamma\leq S$. Applying the YZA relation
leads to \begin{equation}
E_{v}\left[\rho_{{\rm fs}}(S,\gamma)\right]=E_{v}\left[\rho(S,S))\right]=E_{v}\left[\rho(S,-S)\right].\label{eq:FracE2}\end{equation}
As in the case of $S=\frac{1}{2}$ where the fractional spin state
$\rho_{{\rm fs}}(\frac{1}{2},0)$ describes the dissociation limit of a
single chemical bond, the fractional spin state $\rho_{{\rm fs}}(S,0)$
describes the dissociation limit of a multiple chemical bond. This
is demonstrated in Fig. 2 for the dissociation of a double bond, C$_{2}$,
a triple bond, N$_{2}$, and a sextuple bond, Cr$_{2}$, into two
$S=1$, $S=\frac{3}{2}$ and $S=3$ fractional spin atoms respectively. For the molecules
we perform spin-restricted KS calculations and show the binding curve
with respect to the spin-unrestricted ground-state atoms calculated
with no fractional spin ($m_s=S$). On the right hand side of Fig. 2
we show spin-unrestricted calculations on the atoms with fractional spins,
also relative to the normal ($m_s=S$) spin-unrestricted atoms. The expression
for the density, $\rho_{{\rm fs}}(S,\gamma)$, is exactly the same as Eq.
(\ref{eq:KSdensity}) but with fractional occupation, $n_{\rm HOMO}^{\alpha}=\frac{1}{2}+\frac{\gamma}{2S}$
and $n_{HOMO}^{\beta}=\frac{1}{2}-\frac{\gamma}{2S}$, for the top 2$S$ multiple HOMOs
(e.g. the N atom, $\rho_{{\rm fs}}(\frac{3}{2},0)$ has half an $\alpha$
electron and half a $\beta$ electron in the top three $2p$ orbitals).
To compare to molecular dissociation the two densities mixed in Eq.
(\ref{eq:FracE2}) must have the same symmetry. All DFAs
violate the constancy condition, Eq. (\ref{eq:FracE2}). The over-estimation
in the energy for molecular dissociation matches exactly the over-estimation
for the dissociating atoms with fractional-spin density $\rho_{{\rm fs}}^{\sigma}(S,0)$.

The error in the energy for molecular dissociation is normally attributed
to the lack of static correlation, which is remarkably captured by
the violation of the constancy condition for the fractional spin states
of the dissociating atoms. It is thus natural to define a quantitative
measure of the static correlation error (SCE) for approximate density
functionals as \begin{equation}
\textrm{SCE}=E_{v}\left[\rho_{{\rm fs}}(S,0)\right]-E_{v}\left[\rho(S,S))\right].\label{eq:SCE}\end{equation}
Static correlation can be described with the use of a few Slater
determinants for small molecules. However, for large
and bulk systems, this becomes impractical. It is now clear that SCE is
an inconsistency in the commonly used DFAs. 

The errors are massive and increase with the number of bonds. 
It is also very significant to see that the error at the dissociation limit 
can already dominate the behavior close to the bonding region,
making the limiting behavior analysis of $E[\rho_{fs}[S,0]$ broadly relevant.
For Cr$_2$, SCE make HF and B3LYP fail to describe bound molecules and LDA has a very small range of bonding.
Note that these cases are not only challenges for DFT but also for single
reference wave-function methods. The cases considered here
are homonuclear diatomics but the same arguments apply to the dissociation
of heteronuclear diatomics and more complicated molecules.


For the fractional-spin states $\rho_{{\rm fs}}(S,\gamma)$, we have only
explored the consequence of the two-state ensemble which leads to an understanding
of static correlation. There are, however, more general fractional-spin states:
\begin{equation}
\rho_{{\rm fs}}(S,\left\{ C_{m_{s}}\right\} )=\sum_{m_{s}=-S}^{S}C_{m_{s}}\rho(S,m_{s}),\label{eq:GeneralFracSpin}\end{equation}
 where $0\leq C_{m_{s}}\leq1$ and $\sum_{m_{s}=-S}^{S}C_{m_{s}}=1.$
Based on the YZA relation the fractional
spin constancy relation is \begin{equation}
E_{v}\left[\rho_{{\rm fs}}(S,\left\{ C_{m_{s}}\right\} )\right]=E_{v}\left[\rho(S,m_{s}))\right],\label{eq:FracE3}\end{equation}
 which will also have important consequence for molecules and solids.
What may hinder the exploration of Eq. (\ref{eq:GeneralFracSpin})
is the difficulty with which DFT deals with $\rho(S,m_{s})$ for $\left|m_{s}\right|<S$.
Usually only the state $\rho(S,S)$ is calculated, as it can be constructed
easily from a KS determinant. It is difficult, in general, to construct
a KS determinant for other states $\rho(S,m_{s})$ with $\left|m_{s}\right|<S$.

\begin{figure}[!t]
\includegraphics[width=0.5\textwidth]{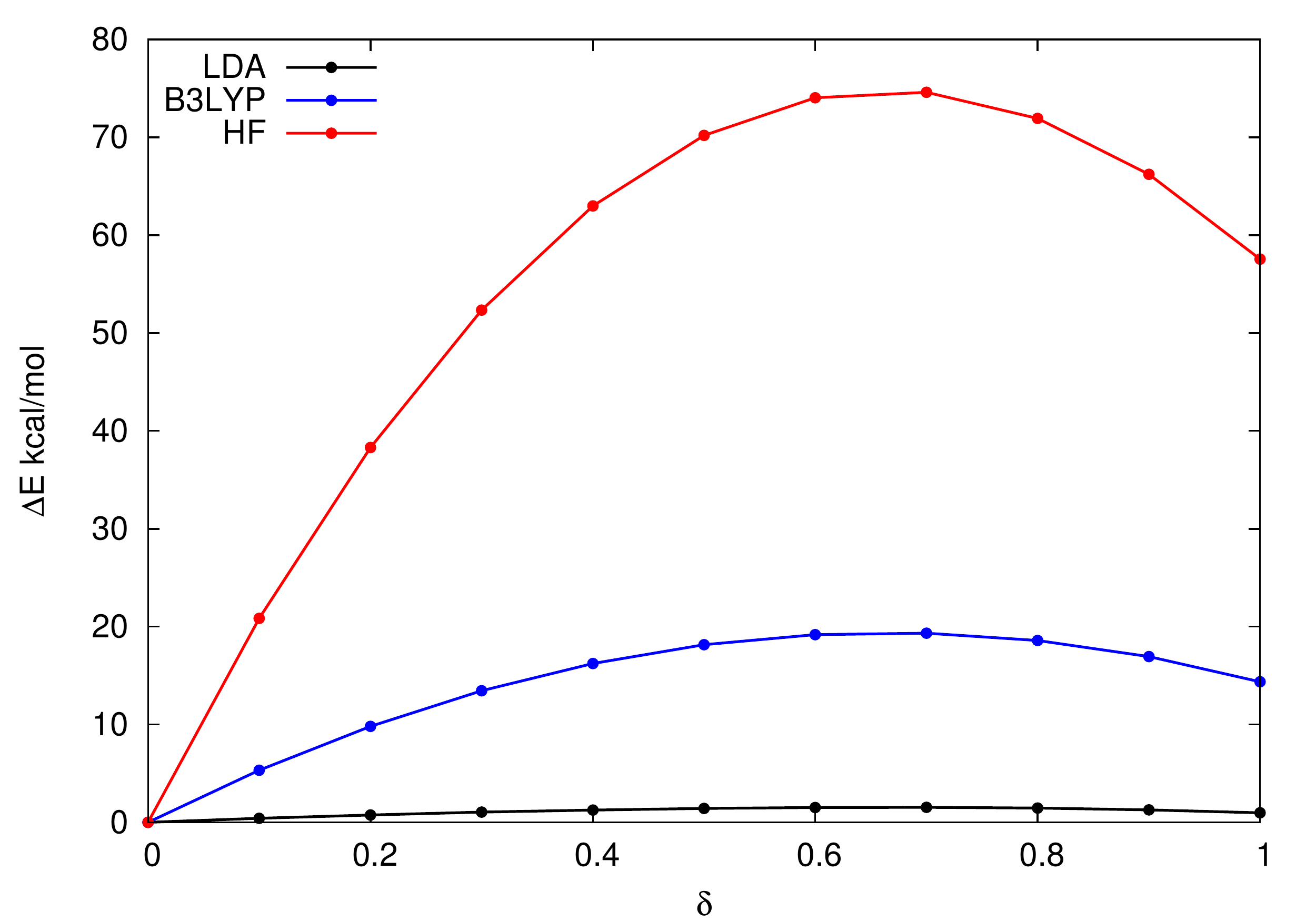} 
\caption{\label{fig3} B atom with fractional alpha spins, $\delta=\frac{2}{3}$
corresponds to the spherical B atom.}
\end{figure}

It is also possible to have fractional-spin states arising from an
ensemble of states which are degenerate because of other symmetries
(e.g. spatial): 
\begin{equation}
\rho_{{\rm fs}}(S,\left\{ C_{i,m_{s}}\right\} )=\sum_{i=1}^{g}\sum_{m_{s}=-S}^{S}C_{i,m_{s}}\rho_{i}(S,m_{s}),\label{eq:GeneralFracSpin2}\end{equation}
 for a $g$-fold degenerate system. The exact energy functional
gives constant energy for all $\left\{ C_{i,m_{s}}\right\} $. For
example a spherical atom density given by $\sum_{i=-L}^{L}\frac{1}{2L+1}\rho_{i}(S,S)$
has alpha fractional spin occupation of the spatially degenerate states.
If we consider the case of the B atom, which is three-fold degenerate,
the lowest energy state predicted by DFAs is given by a non-spherical
density. We now examine the energy of 
$\rho_{{\rm fs}}(S,\{C_\delta\})=\frac{\delta}{2}\left[\rho_{1}(S,S)+\rho_{2}(S,S)\right]+\left(1-\delta\right)\rho_{3}(S,S)$
 such that at $\rho_{{\rm fs}}(S,\{C_0\})$ corresponds to the normal ground state
non-spherical atom and $\rho_{{\rm fs}}(S,\{C_\frac{2}{3}\})$ corresponds
to the spherical B atom. The energy of the fractional spin states
relative to the ground state of the B atom is plotted in Fig. 3, and
shows again the violation of the constancy relation for DFAs.
The error of HF is similar to the case for $\rho_{{\rm fs}}(S,\gamma)$
but with pure DFT the error is much smaller (only $\approx1-2$ kcal/mol
for LDA).

Eq. (\ref{Theorem}) is also valid for energy functionals of the
first-order reduced density matrix \cite{Yang005172}, therefore, our discussion and main results,
Eqs. (\ref{eq:FracE2}) and (\ref{eq:FracE3}), hold in reduced density-matrix functional theory.

In summary this Letter highlights a basic error of DFAs
for degeneracy problems, that are also applicable to the case of near degeneracy. 
These situations can be described within DFT by fractional spin states that are ensembles of degenerate pure-spin
states. This is a simpler concept than the traditional multi-configurational
view, which places any solution outside the realm of normal KS DFT.
It is now clear that the error is solely in the exchange-correlation
functional. The exact constancy relation for the energy derived in
this Letter quantifies the SCE of functionals and shows the basic
error which needs to be addressed for the proper description of strongly
correlated systems, such as magnetic molecules and solids, band structures
of superconductors and Mott insulators. Satisfying
a similar straight line condition for fractional charges has been
very important in addressing the DE of functionals \cite{Cohen07191109}, and we expect the
exact condition of constancy of $E[\rho_{{\rm fs}}]$ to offer a new challenge
for $E_{xc}[\rho]$ and open new frontiers of DFT for strongly correlated systems.

This work has been supported by the National Science Foundation.

\appendix
\section{Fractional spin variational principle and its generalization}
Our starting point is the exact result for an ensemble of degenerate
densities derived by Yang, Zhang and Ayers (YZA) \cite{Yang005172}:
For a $N$-electron system in the external potential $v(\mathbf{r})$
that has $g$-fold degenerate orthogonal ground state wavefunctions
$(\Phi_{i},i=1,2,\ldots,g)$ with corresponding densities $(\rho_{i},i=1,2,\ldots,g)$
and ground state energy $E_{v}^{0}(N)$, the ensemble density is
\begin{equation}
\rho=\sum_{i=1}^{g}C_{i}\rho_{i},\label{Aeq:EnsembleRho}\end{equation}
where $0\leq C_{i}\leq1$ and $\sum_{i=1}^{g}C_{i}=1.$ The exact
energy functional satisfies the following equation \begin{equation}
E_{v}\left[\sum_{i=1}^{g}C_{i}\rho_{i}\right]=E_{v}\left[\rho_{i}\right]=E_{v}^{0}(N),\label{ATheorem}\end{equation}
 if $E_{v}^{0}(N)\leq(E_{v}^{0}(N+1)+E_{v}^{0}(N-1))/2.$

In deriving Eq. (\ref{ATheorem}), YZA used only pure states and the
three requirements on the density functional: (1) correct for each
degenerate orthogonal ground state $(\rho_{i},i=1,2,\ldots,g)$, (2)
translationally invariant and (3) size-consistent. The ensemble densities
thus appear in the limit of large separation of fragments \cite{Yang005172}.
In this sense, we see that with density functional theory (DFT) and
density matrix functional theory, we are forced to define functionals
of ensemble densities.

To construct the functional of ensemble densities of the type of Eq.
(\ref{Aeq:EnsembleRho}), we consider the following trial ensemble
density matrix
\begin{equation}
\tilde{\Gamma}=\sum_{i=1}^{g}C_{i}\tilde{\Gamma}_{i},\label{Aeq:trialDenstiyMatrix}\end{equation}
where $0\leq C_{i}\leq1$ and $\sum_{i=1}^{g}C_{i}=1.$ The density
matrix $\tilde{\Gamma_{i}}$ is for a pure state corresponding to
$i$th degenerate state (in terms of spin, symmetry), but not neccessarily
the ground state. Then we can use the constrained search \cite{Levy796062,Valone804653}
and define the functional for the trial ensemble density $\tilde{\rho}$ as
\begin{equation}
F[\tilde{\rho}]=\min_{\tilde{\Gamma}\rightarrow\tilde{\rho}}{\normalcolor {\normalcolor }\mathrm{Tr\left(\tilde{\Gamma}(\hat{T}+\hat{V}_{ee})\right)}}.\label{Aeq:EnsembleF}\end{equation}
The total energy functional is then
\begin{equation}
E_{v}[\tilde{\rho}]=F[\tilde{\rho}]+\int\tilde{\rho(\mathbf{r})}v(\mathbf{r})d\mathbf{r}.\label{Aeq:EnsembleE}\end{equation}
The minimum of $E_{v}[\tilde{\rho}]$ is the ground state energy
\begin{equation}
E_{v}^{0}(N)=\min_{\tilde{\rho}}E_{v}[\tilde{\rho}],\label{Aeq:VariationalP}\end{equation}
independent of the mixing coefficients $\left\{ C_{i}\right\} $,
because the trial ensemble density matrix, Eq.
(\ref{Aeq:trialDenstiyMatrix}), cannot have an energy lower than the
ground state energy. This is the general variational principle used
in the text for the particular fractional spin systems.

While it may appear as a direct consequence of ensemble DFT, the
variational principle of Eq. (\ref{Aeq:VariationalP}), unlike a
general ensemble DFT theory, connects directly to normal pure-state
DFT calculations without ensembles. The key is the following: the
particular trial ensemble density of Eq. (\ref{Aeq:EnsembleRho}),
which consists of densities from orthogonal degenerate ground
states, arises directly as the dissociation limit of normal pure
systems  \cite{Yang005172}. Such pure states are calculated with
normal pure-state  DFT with one KS determinant, making the YZA
analysis and our variational principle of Eq.
(\ref{Aeq:VariationalP}) directly relevant.


In carrying out corresponding self-consistent KS calculations, we
have also assumed that such trial ensemble densities $\tilde{\rho}$
can be represented by a noninteracting systems with fractional occupation.
This parallels the development in the self-consistent KS calculations
of fractional charge systems \cite{Morisanchez06201102}
where the fractional charge ensemble is also represented by a noninteracting
system with fractional occupation.

While the validity of such a KS representation has not been mathematically
established here, just as in the case of the KS representation for
normal pure state densities, the most important justification is that
such fractional occupation KS calculations reveal important features
in the functionals for fractional spin ensembles as reported in this
work and fractional charge ensembles as reported earlier \cite{Morisanchez06201102}. These
important failures of density functionals are observed in normal
DFT calculations without using fractional charge nor spin. When they
are analyzed in terms of the fractional occupation KS calculations,
they become transparent as a clear violation of the fundamental
equalities: linearity in the fractional charge case and constancy
in the fractional spin case.

\begin{figure}[!t]
\includegraphics[width=0.5\textwidth]{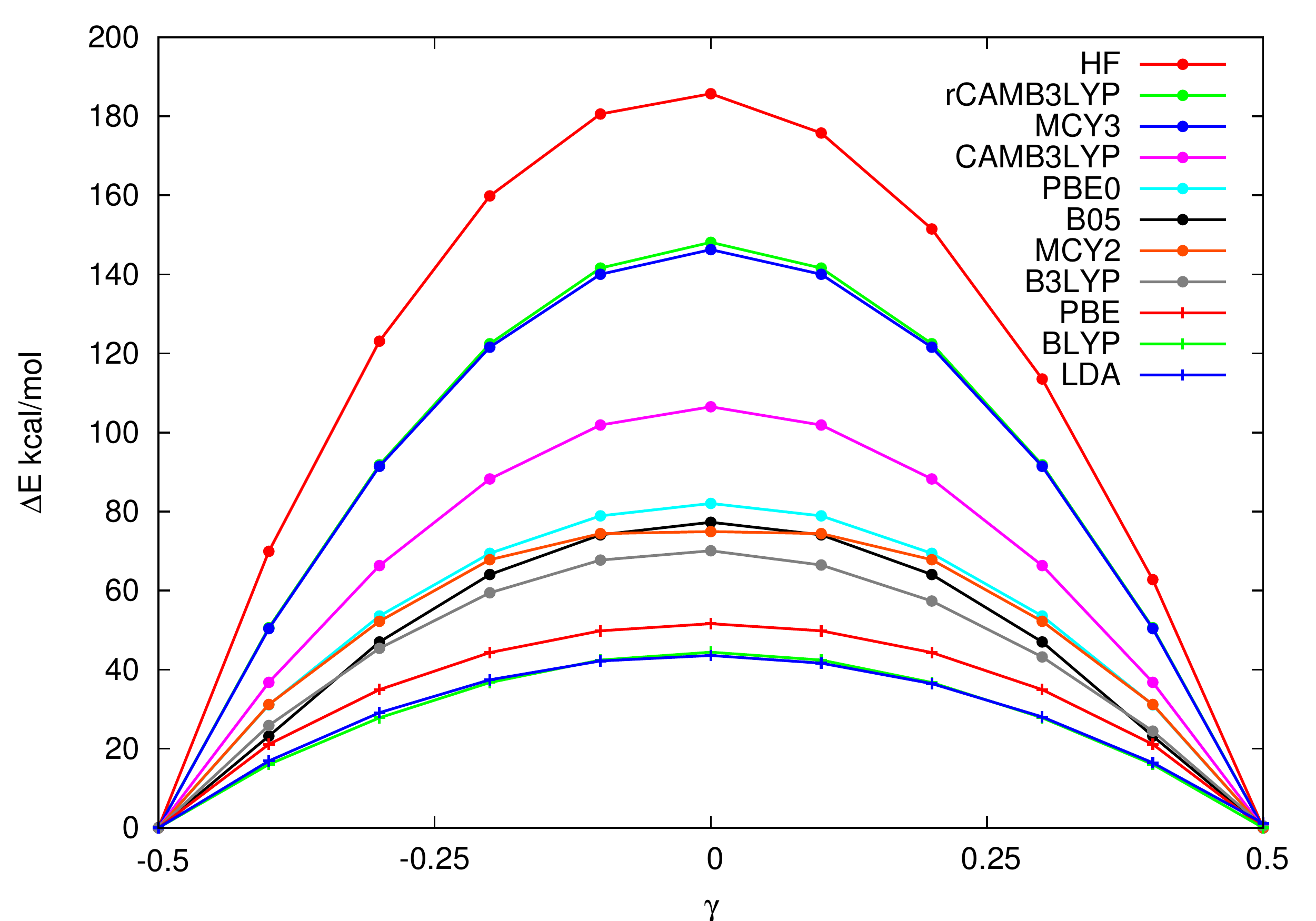}
\caption{\label{figA} Spin-unrestricted KS calculations, using several different functionals,
on the H atom with fractional spins (multiplied by 2). $\gamma=0$ is a H atom with half an $\alpha$
electron and half a $\beta$ electron, which is the dissocation limit
of H$_{2}$. All calculations are self-consistent using a cc-pVQZ basis set.}
\end{figure}

\section{Results for the H atom}

The performance of several different functionals is shown in Fig. 1 for the fractional spin H atom.
The smallest error is seen for LDA and GGA functionals \cite{Becke883098,Lee88785,Perdew963865} 
with hybrid functionals B3LYP \cite{Becke935648}, MCY2 \cite{Mori-Sanchez06091102}, B05 \cite{Becke05064101} and PBE0\cite{Adamo996158}
all having very similar performance despite their varied forms. Functionals with improved behaviour
on delocalization error, such as MCY3 and rCAMB3LYP \cite{Cohen07191109}, which have a much straighter line for fractional
charges, have a very poor performance for this fractional spin problem. Note also that functionals
which are exact for the integer spin H atom such as HF, MCY2 and B05 have large errors for fractional spin.

\end{document}